\documentclass[3p,twocolumn,authoryear]{elsarticle}

\usepackage{amssymb}
\usepackage{amsmath}
\usepackage[colorlinks,citecolor=blue]{hyperref}
\usepackage{lineno}
\usepackage{booktabs} 
\usepackage{ulem}

\journal{Journal of High Energy Astrophysics}
\begin{document}

\begin{frontmatter}

\title{The late flare in tidal disruption events due to the interaction of disk wind with dusty torus}

\author[a1]{Jialun Zhuang}
\ead{zhuangjlun@mail2.sysu.edu.cn}
\author[a1]{Rong-Feng Shen}
\ead{shenrf3@mail.sysu.edu.cn}
\affiliation[a1]{organization={School of Physics and Astronomy},
            addressline={Sun Yat-Sen University}, 
            city={Zhuhai}, 
            postcode={519000},
            state={Guangdong},
            country={People's Republic of China}}

\begin{abstract}
A late (t $\sim$ 1,500 days) multi-wavelength (UV, optical, IR, and X-ray) flare was found in PS1-10adi, a tidal disruption event (TDE) candidate that took place in an active galactic nucleus (AGN).
TDEs usually involve super-Eddington accretion, which drives fast mass outflow (disk wind). 
So here we explore a possible scenario that such a flare might be produced by the interaction of the disk wind with a dusty torus for TDEs in AGN. 
Due to the high velocity of the disk wind, strong shocks will emerge and convert the bulk of the kinetic energy of the disk wind to radiation.  
We calculate the dynamics and then predict the associated radiation signatures, taking into account the widths of the wind and torus.
We compare our model with the bolometric light curve of the late flare in PS1-10adi constructed from observations.  
We find from our modeling that the disk wind has a total kinetic energy of about $10^{51}$ erg  and a velocity of 0.1 c (i.e., a mass of 0.3 $M_{\odot}$);
the gas number density of the  clouds in the torus is $3\times 10^{7}$ $\rm cm^{-3}$. Observation of such  a late flare can be an evidence of the disk wind in TDEs and can be used as a tool to explore the nuclear environment of the host. 
\end{abstract}


\end{frontmatter}

\section{Introduction} \label{sec:intro}

An active galactic nucleus (AGN) is thought to be powered by a massive accreting black hole residing at the center of the host galaxy \citep{2015ARA&A..53..365N}.
In the unified model of AGN, an axisymmetric dusty structure  composed of dusty gas clumps is one of the essential ingredients \citep{1988ApJ...329..702K, 1993ARA&A..31..473A} and observations also reveal that there are lots of clumpy molecular clouds in the host central region \citep{1996A&ARv...7..289M, 2008ApJ...678L..13K}. The density of these gaseous clumps lies in a wide range of $n_{H} \sim (10^{6}-10^{8}) $ $\rm cm^{-3}$ \citep{1993ApJ...402..173J, 2005ApJ...622..346C}.

With a super massive black hole (SMBH $10^{6} - 10^{9} M_{\odot}$) at the center, stars in a galaxy can occasionally be scattered to approach the SMBH so close that they can be tidally disrupted, known as a tidal disruption event (TDE) \citep{1988Natur.333..523R, 1989IAUS..136..543P}. 
Almost half of the debris would escape while the rest of the stellar mass would be accreted into the black hole leading to a luminous flare lasting for months to years with luminosity decaying as $t^{-5/3}$ \citep{1988Natur.333..523R, 1989IAUS..136..543P}.

A reasonable fraction of the matter initially bound to the black hole is likely to be blown away by the huge radiation pressure when the fallback rate of these bound debris is super-Eddington at early times \citep{2009MNRAS.400.2070S}.
For black holes with masses greater than  $10^{7} M_{\odot}$ whose tidal disruption radius is highly relativistic, the self-crossing of the fallback stream due to general relativistic precession  can also generate energetic, fast outflow \citep{2016ApJ...830..125J, 2020MNRAS.492..686L}.
The typical velocity of the disk wind is $ 0.1$ c where c is the speed of light \citep{2017ApJ...843L..19M, 2018ApJ...859L..20D}. 
For example, in the simulation of \citet{2016MNRAS.458.4250S}, about $10 \%$ of the star’s total mass can be transferred into outflow with a velocity of $\sim $ 0.1 c.
Outflows with high speeds up to 0.1 c have also been found in observations of a TDE \citep{2017ApJ...843..106B}.

TDEs happening in AGN used to be neglected for its lower observational prospect due to the likely contamination from the AGN variability. 
With the development of transient surveys, it has been taken attention.
Recently, \cite{2017NatAs...1..865K} reported 
an optical transient named PS1-10adi which occurred in August 2010. It is located in the nucleus of the galaxy SDSS J204244.74+153032.1 at a distance of d=1.00$\pm$0.01 Gpc (z=0.203$\pm$0.001).
The central black hole mass is estimated to be  $M_{BH}$ =$2.7\times 10^{7}$ $M_{\odot}$.  
Unlike any known types of recurring AGN variability, the light curve of PS1-10adi  evolves slowly and smoothly for 1,000 days with a total radiated energy of $2.3\times 10^{52}$ erg. Furthermore, this transient reached its peak luminosity of $10^{44}$ erg $\rm s^{-1}$ soon after the first detection and declined exponentially.

Although PS1-10adi occurred in the AGN central region, \cite{2017NatAs...1..865K} suggested that it was not associated with the host AGN activity because known classes of the latter could not account for this transient. 
\cite{2019ApJ...871...15J} presented the infrared counterpart of PS1-10adi and interpreted it as the re-emission of a dusty torus as it was heated by the UV emission from the central TDE, i.e., the dust echo of a TDE which was first predicted by \cite{2016MNRAS.458..575L}. 

\cite{2017NatAs...1..865K} have also presented UV-optical-IR light curves of PS1-10adi up to $\sim$ 2,200 days since the peak, in which a secondary bump (a factor of $\sim10$ increase) stands out at about 1,500 days. 
Hereafter we call this late bump a rebrightening or a late flare. 
The luminosity of this bump rose rapidly in about 100 days and then reached its peak luminosity of 
about $10^{43}$ erg $\rm s^{-1}$. It shows a sharp decline as well in about 100 days after the peak.
It was detected in X-rays right before the UV-optical peak with a 0.3-10 keV luminosity of $10^{43}$ erg $\rm s^{-1}$, although it lacks X-ray observation at later times.

Shocks are abundant in the universe and they are generated when obstacles are exposed to fast flows or, vice versa \citep{2009A&ARv..17..409T}. 
As the shock propagates, it transforms the kinetic energy of the fast flow into thermal energy of the shocked medium and eventually radiates as the shocked hot matter cools.
For TDEs in AGN, it is natural that disk wind may collide with the torus, leading to the formation of shocks and then producing a late flare as was seen in PS1-10adi.
This scenario has been proposed in \cite{2019ApJ...871...15J} and we investigate it carefully here.

Similar scenarios in other contexts have been studied in \cite{2016ApJ...822...48G} and \cite{2017ApJ...843L..19M}.
 When our work was finished and this manuscript was being prepared, we notice that this scenario of interaction of wind with torus has been discussed also by \cite{2020arXiv200910420M}. 

This paper is organized as follows. 
In Section 2 we construct an analytical model for the hydrodynamical evolution of the disk wind-torus interaction.
In Section 3 we consider the radiation processes and calculate the light curves.
We apply the light curve model to the rebrightening of PS1-10adi and summarize the results in Section 4. 
The conclusion and discussion are given in Section 5.

\section{Hydrodynamics of the disk wind-torus interaction} \label{sec:analysis}
Though there are discussions about how the disk wind is generated and distributed, our model is independent of the wind launching mechanism.
Since the torus is only distributed in the equatorial plane, here we assume that  the solid angle spanned by the torus is $\Omega$ and the wind is isotropic. We only pay attention to the angular portion of the wind that would collide with the torus. 
\begin{figure}[htbp]
\centering
\includegraphics[width=8cm]{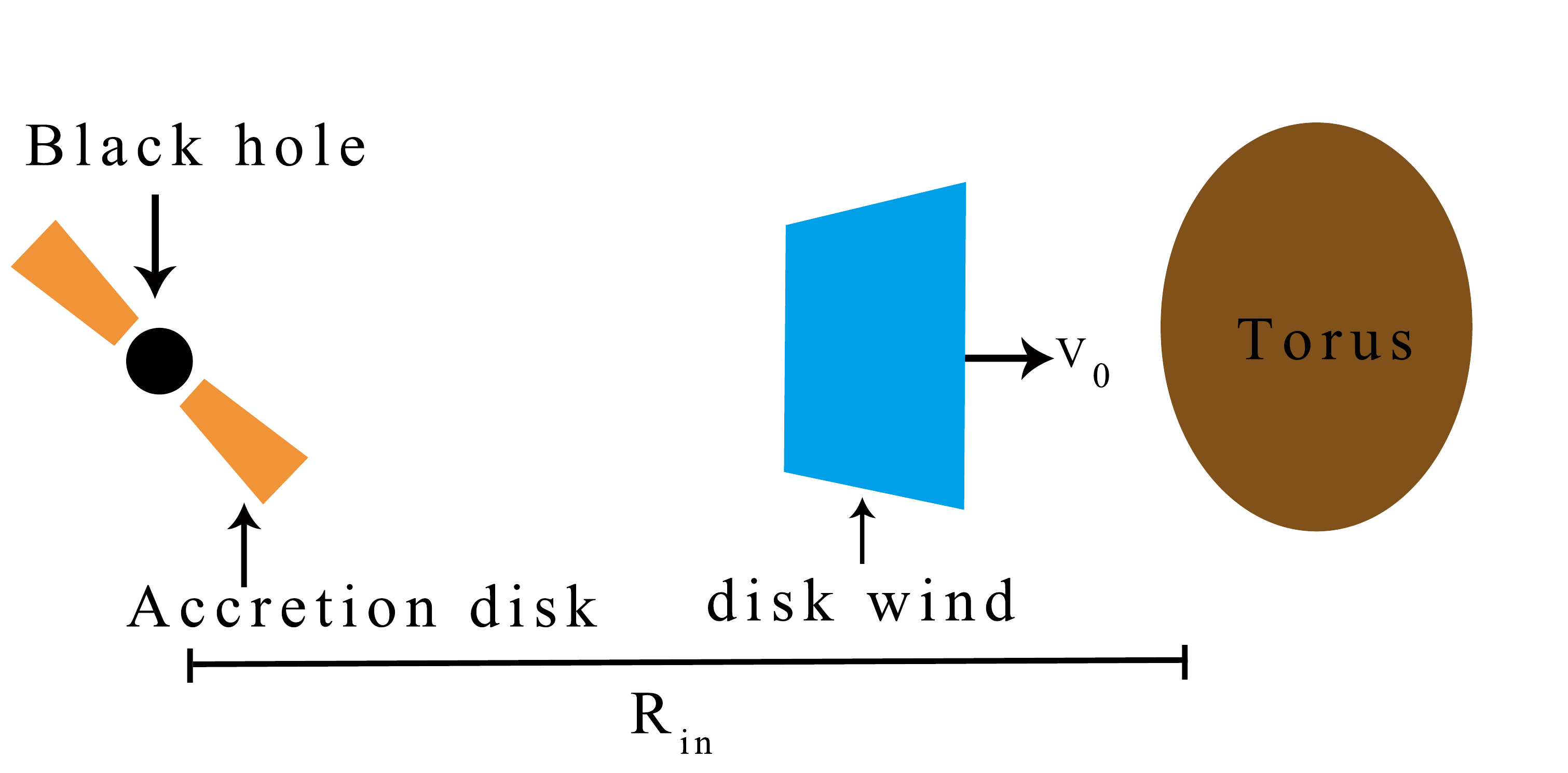}
\caption{The schematic of the disk wind-torus interaction model (not to scale). 
The distance between the black hole and the torus is $R_{in}$. The brown region is the uniform torus. The disk wind expands freely with velocity $v_{0}$  before colliding with torus and this collision would occur at $t_{in}=R_{in}/v_{0}$ after the tidal disruption event. \label{fig:structure}}
\end{figure}

We take the radius of these clumpy clouds in the torus as $R_{c}=0.01$ pc.
According to the observation of PS1-10adi, we can compare the size of the cloud with the depth that disk wind would propagate within a few hundred days  by 
assuming that the disk wind velocity is 0.1 c.
Thus, we conclude that the molecular cloud is thick enough to stop the colliding disk wind, which means that the colliding disk wind will end up mixing with the cloud and will not interact with another cloud. Therefore, we consider that  the torus is uniform 
with a number density $n_{0}$ same as that of the molecular cloud.
Also, since the evolution timescale of the late flare in PS1-10adi is much smaller than $t_{in}$,
 the depth of the forward shock into the torus $D$ is much smaller than the inner radius of torus $R_{in}$ as well.

Here we consider the collision between the disk wind and the torus,  both of which are homogenous with a finite width, $W_{w}$ and $W_{t}$ respectively. This collision is illustrated in Figure \ref{fig:structure}.
The interaction between the wind and the cloud would lead to the formation of a forward shock that sweeps up the  gas clouds in the torus,  and a reverse shock that runs into the wind \citep{1999ApJS..120..299T, 2019MNRAS.488..978J}.

\begin{figure}[h]
\centering
\includegraphics[width=7.5cm]{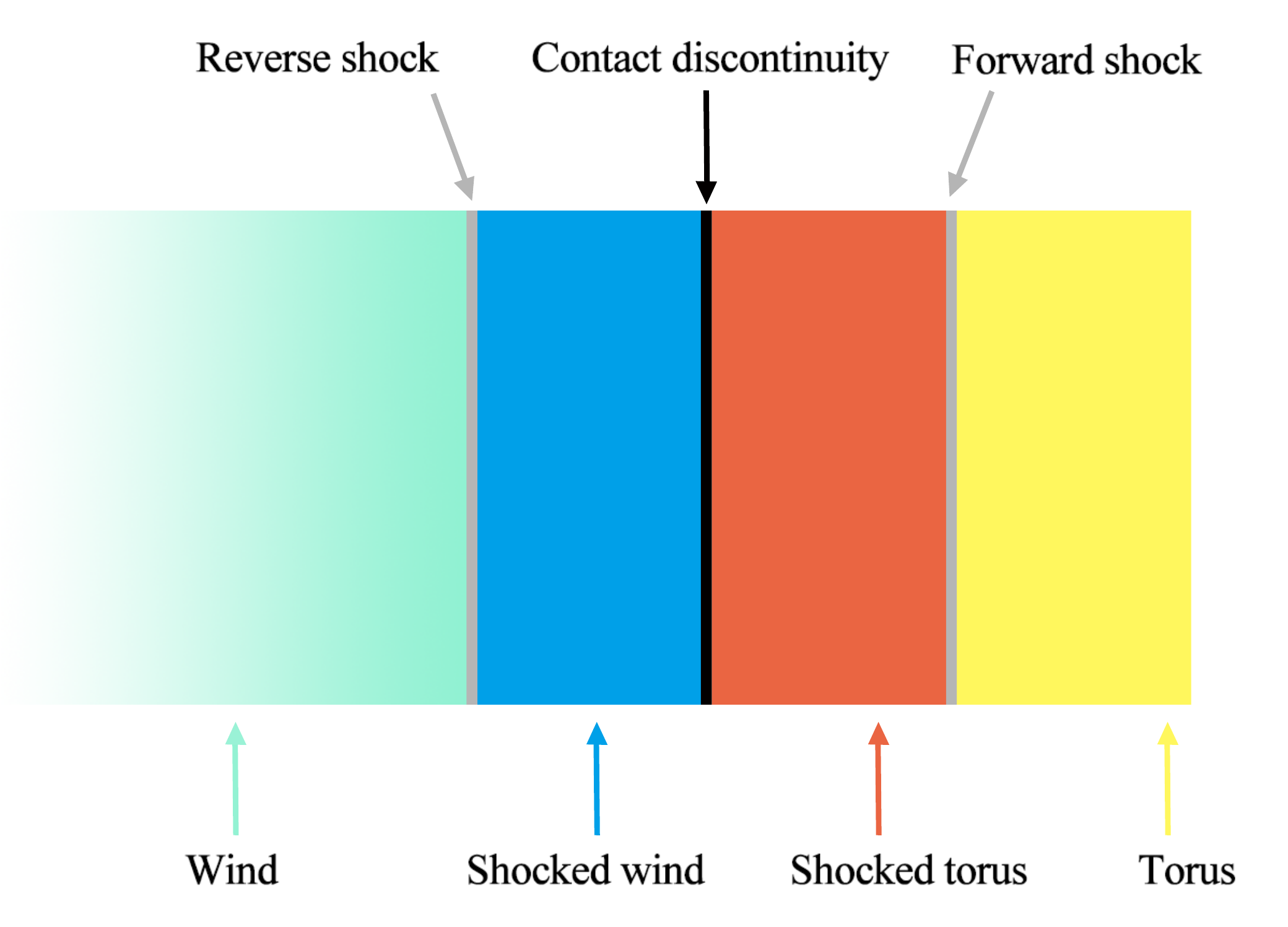}
\caption{Schematic structure after the disk wind-torus collision.\label{fig:structure2}}
\end{figure}

Thus for convenience, we divide the structure into four regions, the unshocked torus-A, the shocked torus-B, the shocked wind-C, and the free wind-D as shown in Figure \ref{fig:structure2}. 
The subscripts $A$, $B$, $C$, $D$ just refer to the four regions mentioned above, and subscripts $FS$, $RS$ refer to the forward shock and reverse shock respectively;
 $t$ is the time starting at the moment of collision;
$t_{FS}$  is the time for the forward shock to go across the whole torus while $t_{RS}$ is the time for the reverse shock to cross the entire wind;
 $E_{0}$ is the total kinetic energy of disk wind;
$M_{ej}$ is the total mass of the colliding disk wind;

Now we can construct the theoretical model and look into the dynamical interaction between disk wind and torus.
At first, we analytically investigate the evolution of the shocks created by the interaction. 
And then, we will provide an analytic expression for the bolometric light curve.

\subsection{Early phase}
Early when the radiative loss is dynamically insignificant and the system is adiabatic, with the Rankine-Hugoniot relations for an adiabatic shock \citep{1959flme.book.....L, 2008ARA&A..46...89R}, we have
\begin{align}
P_{2}&=\frac{2}{\gamma+1} \rho_{1} v_{1}^{2}\\
T_{2}&=\frac{3\mu m_{H}v_{1}^{2}}{16k}\\
v_{2}&=\frac{\gamma-1}{\gamma+1} v_{1}\\
\rho_{2}&=\frac{\gamma+1}{\gamma-1}\rho_{1}
\end{align}
where pre-shock quantities such as density and pressure are labeled with a subscript 1, and post-shock quantities with a subscript 2; $v_{1}$, $v_{2}$ are the velocities relative to the shock; $\gamma=5/3$ is the adiabatic index,  $\mu m_{H}$ is the is mean mass per particle and $\mu$ is
set to be 13/21, appropriate for solar abundances \citep{1998ApJ...500..342B}.

Several calculations \citep{ 1973MNRAS.161...47G, 1975MNRAS.171..263G, 1982ApJ...259..302C, 1984ApJ...281..682H, 2018MNRAS.478.5112S} have shown that the thermal pressure of the shocked ejecta  and the shocked ambient gas rapidly becomes almost homogeneous which gives $P_{B}=P_{C}$. 
Since the shocked wind moves with the same velocity as the shocked cloud,  we have
\begin{align}
v_{B}&=v_{D}\frac{1}{\alpha+1}\\
v_{FS}&=\frac{\gamma+1}{2} v_{B}\\
v_{RS,D}&=\frac{\gamma+1}{2} (v_{B}-v_{D}) \label{eqn:vrsd}
\end{align}
where $v_{RS, D}$ is the reverse shock velocity in the frame of the unshocked wind; $\alpha= (\rho_{A}/\rho_{D})^{1/2}$ and $\alpha >1$ is concerned.
These velocities remain nearly unchanged  in the early phase as long as the densities upstream stays the same.

The energy in these shocked regions are 
\begin{align}
E_{B}
&\approx M_{B}v_{B}^{2}\\
E_{C}
& \approx (\alpha^{2}+1)\frac{1}{2}M_{C}v_{B}^{2}
\end{align}
and the depth and velocity of the forward shock are
\begin{align}
D_{FS}(t)&=v_{FS}t\\
v_{FS}(t)&=v_{FS}
\end{align}

\subsection{The cooling timescale $t_{cool}$}
As is known, the shocked matter would cool eventually and the shock can't be considered adiabatic later. This transition is marked by a catastrophic collapse of the postshock gas, forming a thin, dense shell behind the shock.

\begin{figure}[htbp]
\centering
\includegraphics[width=8cm]{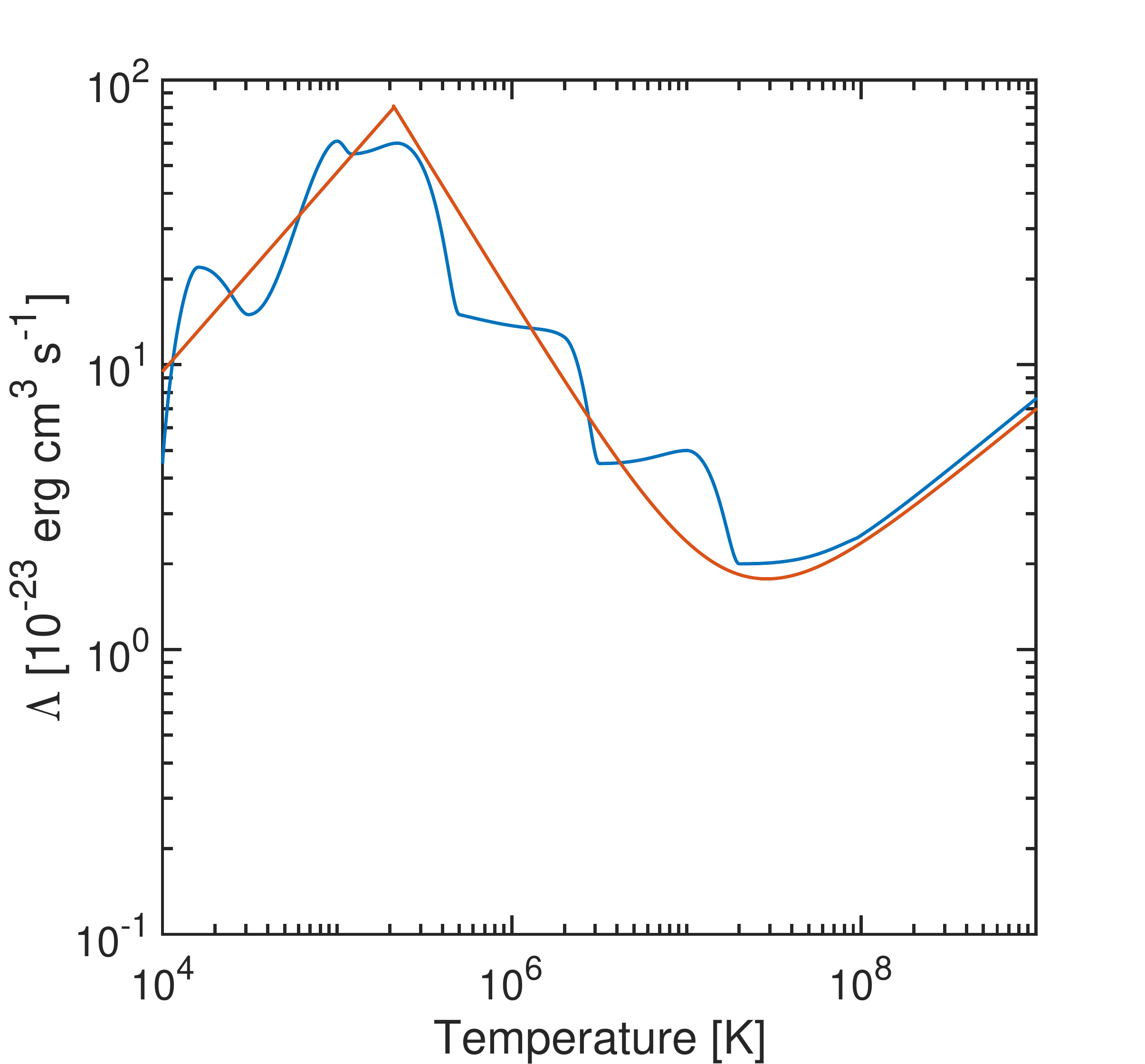}
\caption{ The piecewise power law fit (red)  to the total cooling coefficient (blue) of \cite{1976ApJ...204..290R}. \label{fig:cooling}}
\end{figure}

One can obtain a crude estimate of the age at which the shocked matter becomes radiative by comparing the cooling time of gas immediately behind the shock with the age of the evolution,
 and the cooling time is \citep{1998ApJ...500..342B}:
\begin{equation} \label{eqn:tcool}
  t_{cool}=\frac{0.69kT_{2}}{n_{1}\Lambda(T_{2})},
\end{equation}
where $n_{1}$ is the number density upstream, $k$ is the Boltzmann constant and $\Lambda(T)$ is the temperature-dependent volume cooling function
\citep{1971ApJ...167..113C, 1979ApJ...227..131S}.  
$\Lambda$ is in units of erg $\rm cm^{3}$ $\rm s^{-1}$ and is defined such that $n^{2} \Lambda$ gives the emitted power per unit volume \citep{1988ApJ...334..252C}. 
For $\Lambda (T)$ we use the following piecewise power-law fit to the  numerical result of \cite{1976ApJ...204..290R}  as shown in Figure \ref{fig:cooling}:
\begin{align} \label{eqn:lambda}
\Lambda \approx
\begin{cases}
1.5\times10^{-25} T^{0.7},		&{T \le 2\times10^{5}\  \rm K }\\
\ \\
1.7\times10^{-16}T^{-1}+\\
\ \ \ 2.2\times10^{-27}T^{\frac{1}{2}}, &T>2\times10^{5}\  \rm K
\end{cases} 
\end{align}
The term $2.2\times10^{-27}T^{\frac{1}{2}}$ comes from the  bremsstrahlung emission \citep{1961ApJS....6..167K, 1969ApJ...157.1157C} and the term $1.7\times10^{-16}T^{-1}$ is adopted   according to \cite{1998ApJ...500..342B}.

When the temperature of the shocked regions drops below $10^{8}$ K, we approximate $\Lambda =10^{-16}T^{-1}$  in the range of $10^{5}$ K$<T<10^{8}$ K to estimate the age of transition for the sake of the simplicity.
Since cooling has affected the evolution prior to $t=t_{cool}(t)$, we begin the radiative stage a factor of $e$ sooner where $e<1$ \citep{cioffi_mckee_1988} as $t_{cool}=e\times t_{cool}(t)$.

For $\alpha>1$, we found that $t_{cool}$ of the shocked wind is much greater than that of the shocked torus  due to a lower density. Thus we only pay attention to $t_{cool}$ of the shocked torus and $t_{cool}$ in the following all refer to the $t_{cool}$ of the shocked torus if not specified.

\subsection{Succeeding evolution}
Following further expansion, the shock velocities may change due to a variety of reasons, for example, the reverse shock has passed the whole wind, the forward shock has crossed the cloud or the shocks become radiative.
Therefore, we discuss the subsequent evolution  according to the order of $t_{RS}$, $t_{FS}$ and $t_{cool}$.

\subsubsection{Case 1 - $t_{RS}<t_{cool}\ll t_{FS}$}
 In this case, the width of the torus can be considered infinitely wide so that $t_{FS}$ is much larger than $t_{RS}$ and $t_{cool}$. With equation \ref{eqn:vrsd}, $t_{RS}$ can be estimated as:  
 \begin{align}
t_{RS}&\approx \frac{3W_{w}}{4v_{RS, D}} \label{eqn:trs}
\end{align}

After $t_{RS}$,
the system enters into the Sedov-Taylor phase for the energy of the free wind has transferred to the shocked matters and the radiative loss is still not energetically important at this time \citep{1950RSPSA.201..159T,1959sdmm.book.....S}.
As a rough estimate, one considers that the entire energy of the disk wind is transferred to the shocked torus medium \citep{2012A&ARv..20...49V}.
Applying the conservation of energy, we have
\begin{equation}
E_{0}=  \rho _{A} R_{in}^{2}D(t)\ \left[ \frac{d(D(t))}{dt} \right]^{2}.
\end{equation}

If we try a self-similar solution for the depth as a power-law function of time D(t) $\propto$ $t^{a}$
and substitute it into the equation above, we get $a=2/3$. 
So the forward shock's position and velocity will evolve as
 \begin{align}
 D_{FS}(t)&=D_{FS}(t_{RS})(\frac{t}{t_{RS}})^{\frac{2}{3}}\\
 v_{FS}(t)&=v_{FS}(t_{RS})(\frac{t}{t_{RS}})^{-\frac{1}{3}}
\end{align} 
in this phase.

So far, it was assumed that although the hot gas might lose or gain energy by doing $PdV $ work, the total energy of the shock waves was constant \citep{1999ApJS..120..299T},
so the free expansion and the Sedov-Taylor phase are also classified as adiabatic phases.

The shock enters radiative phase after $t_{cool}$ and is driven by the pressure of the hotter shocked ambient gas deeper in the interior \citep{1980ARA&A..18..219M,1998ApJ...500..342B}. The evolution of the shock position is now best described by using the momentum conservation \citep{2012A&ARv..20...49V}:
 \begin{equation}
M_{B}(t_{cool})v_{FS}(t_{cool})= R^{2}_{in} D(t) \rho_{A} \ \frac{dD(t)}{dt}
\end{equation} 
Same as before, if we try a power-law solution for the depth as the function of time, we have 
 \begin{align}
D_{FS}(t)& =D_{FS}(t_{cool})(\frac{t}{t_{cool}})^{\frac{1}{2}}\\
v_{FS}(t)& =v_{FS}(t_{cool})(\frac{t}{t_{cool}})^{-\frac{1}{2}}
\end{align} 

Throughout the radiative stage, the shocked region loses most of its thermal energy \citep{2019MNRAS.488..978J}. It finally merges with the surrounding medium when its expansion velocity becomes comparable to the sound speed in the ambient gas.

\subsubsection{Case 2 - $t_{cool}<t_{RS}\ll t_{FS}$}
In this case, the disk wind is wide enough that the forward shock becomes radiative before $t_{RS}$. And we can not use the Rankine-Hugoniot relations to calculate the pressure in the shocked cloud region anymore after $t_{cool}$.

By this time, we think that the pressure in the shocked wind is equivalent to the force exerted by the incoming free torus material per unit time per unit area on the thin shell of the shocked torus region which gives $\rho_{A} v_{B}^{2}=P_{C}$.
With a coefficient different from the previous formula, these velocities  can be considered invariable before $t_{RS}$.
And $t_{RS}$ can also be estimated as equation (\ref{eqn:trs}) in this case.

After $t_{RS}$, with momentum conservation,  the forward shock's position and velocity evolve as 
 \begin{align}
D_{FS}(t)& =D_{FS}(t_{RS})(\frac{t}{t_{RS}})^{\frac{1}{2}}\\
 v_{FS} (t)& =v_{FS}(t_{RS})(\frac{t}{t_{RS}})^{-\frac{1}{2}}
 \end{align}

\subsubsection{Case 3 - $t_{cool}<t_{FS}<t_{RS}$}
When $W_{t}$ is so small that the forward shock would dissipate before $t_{RS}$, $t_{FS}$ can be estimated as
\begin{align}
t_{FS}=\frac{W_{t}}{v_{FS}}
\end{align}
In this case, the $W_{w}$ is so wide and the forward shock becomes radiative before $t_{FS}$, we have $t_{cool}<t_{FS}<t_{RS}$ where we label it as case 3. the evolution is the same as the one described in case 2 before $t_{FS}$.

\subsubsection{Case 4 - $t_{FS}<t_{cool}<t_{RS}$}
Same as case 3, if the shock becomes radiative later than $t_{FS}$, then we get $t_{FS}<t_{cool}<t_{RS}$ where we label it as case 4.
For reasons presented in section 3.2, since the luminosity from the shocked torus will not raise afterward,  we no longer pay attention to its subsequent evolution for both case 3 and case 4.

\section{Radiation} \label{sec:Radiation}
As a shock evolves through the stages outlined above, the radiation it emits changes as well, such as thermal X-ray emission from the increasing volume of shock-heated gas and synchrotron emission from a non-thermal electron distribution \citep{2017hsn..book.1981R}.
Radio observation of PS1-10adi only yielded an upper limit during the main peak, but it is worth discussing since it is believed that the radio emission is usually associated with a fast interaction region \citep{1982ApJ...259..302C}.

\subsection{Bolometric luminosity} \label{subsec: Bolometric luminosity}
We now calculate the radiation expected from the interaction and  construct an analytic bolometric light curve  to compare with the observation.

It is suggested that the electrons in the ionized shock wave can reach temperature equilibrium with the ions during the non-radiative expansion phase, somewhat before cooling becomes important \citep{1968supe.book.....S}, so the bolometric luminosity can be calculated from the cooling function $\Lambda (T)$ for gas in collisional ionization equilibrium. 

But that is not the case once the shock becomes radiative because the shocked gas is not in ionization equilibrium anymore \citep{1988ApJ...334..252C}.
Following further expansion, radiative losses are so important that the thermal energy of all newly swept-up material is rapidly radiated \citep{cioffi_mckee_1988, 1988ApJ...334..252C}.
This implies that the luminosity of a radiative shock is proportional to $r^{2}v^{3}$ \citep{1974ApJ...188..501C, 1988RvMP...60....1O}, where $r$ and $v$ are the position (with respect to the wind launching site) and velocity of the shock respectively;
$r^{2}v$ represents the increment in the volume of the newly shocked region per unit time, so the energy of the newly swept-up material is  $\propto r^{2}v\times v^{2}$. 

Therefore we have 
\begin{eqnarray} \label{eqn:Lbol}
L_{bol}=
\begin{cases}
n_{2}^{2} \Lambda (T_{2}) V_{2},  & {t \le t_{cool} }\\
\ \\
\frac{1}{2}\rho _{1} \Omega r^{2}v_{1}^{3}, & {t>t_{cool}}
\end{cases}
\end{eqnarray}
where $V_{2}$ is the volume of the shocked region.
Since Equation (\ref{eqn:Lbol}) is piecewise, we adjust $e$ as is stated in section 2.2 to make sure that the luminosity evolution is continuous at $t_{2}$ and it shows that $e$ is at the order of  magnitude of about order unity.

The preceding equation is only valid in the case where the shock exits. 
As is mentioned, the shock would dissipate once it crosses the entire matter. 
On this occasion,
with no new energy transferring to the shocked region, the shocked region would cool down gradually, along with the temperature and luminosity.  Besides,  free expansion of the shocked region would cause the temperature to drop faster. Since the luminosity evolution is not that important thereafter, we simply consider that it would decay exponentially.

\subsection {Evolution of luminosities from the two shocked regions}

Combining section 2 and section 3.1, we can estimate the luminosities from the two shocked regions in different cases.

\subsubsection{Case 1 - $t_{RS}<t_{cool}\ll t_{FS}$}
Early, both $L_{B}$ and $L_{C}$ will increase as $L \propto t$  because velocities of these shocks remain fixed while the shocked masses are accumulating, and $L_{B}$ is much greater than $L_{C}$ due to a higher density although $E_{B}$ is less than $E_{C}$.
 After $t_{RS}$, $L_{B}$ keeps increasing as $L_{B} \propto t^{1/3}$ because the shocked mass increases even the shock velocity is decreasing and would reach its peak at $t_{cool}$,
while $L_{C}$ declines exponentially as $L_{C} \propto e^{-t}$.
Later $L_{B}$ would drop as $L_{B} \propto t^{-1/2}$ for the shock velocity declines much faster.

\subsubsection{Case 2 - $t_{cool}<t_{RS}\ll t_{FS}$}
Likewise, $L_{B}$ and $L_{C}$ grow as $L \propto t$ before $t_{cool}$.
Later,
$L_{B}$ keeps nearly constant as $L_{B} \propto t^{0}$ because its velocity  and position ($R_{in}+v_{FS}t$) remain nearly the same for  $t<t_{in}$, while $L_{C}$ would keep increase. 
After $t_{RS}$, $L_{B}$ drops as $L_{B} \propto t^{-1/2}$ with velocity declining, and $L_{C}$ decreases exponentially.

\subsubsection{Case 3 - $t_{cool}<t_{FS}<t_{RS}$}

The evolution of $L_{B}$ is the same as in case 2 before $t_{FS}$, and  drops exponentially after, while $L_{C}$ keeps increasing as $L_{C} \propto t$ before $t_{RS}$. 

\subsubsection{Case 4 - $t_{FS}<t_{cool}<t_{RS}$}
After $t_{FS}$, given that the radiation loss is still negligible, the shocked region is adiabatic,  and the temperature of this shock region is thought to remain unchanged, along with its luminosity, until $t_{cool}$. And it would drop exponentially as $L_{B} \propto e^{-t}$ after.

\begin{figure}[htb]
\centering
\includegraphics[width=8cm]{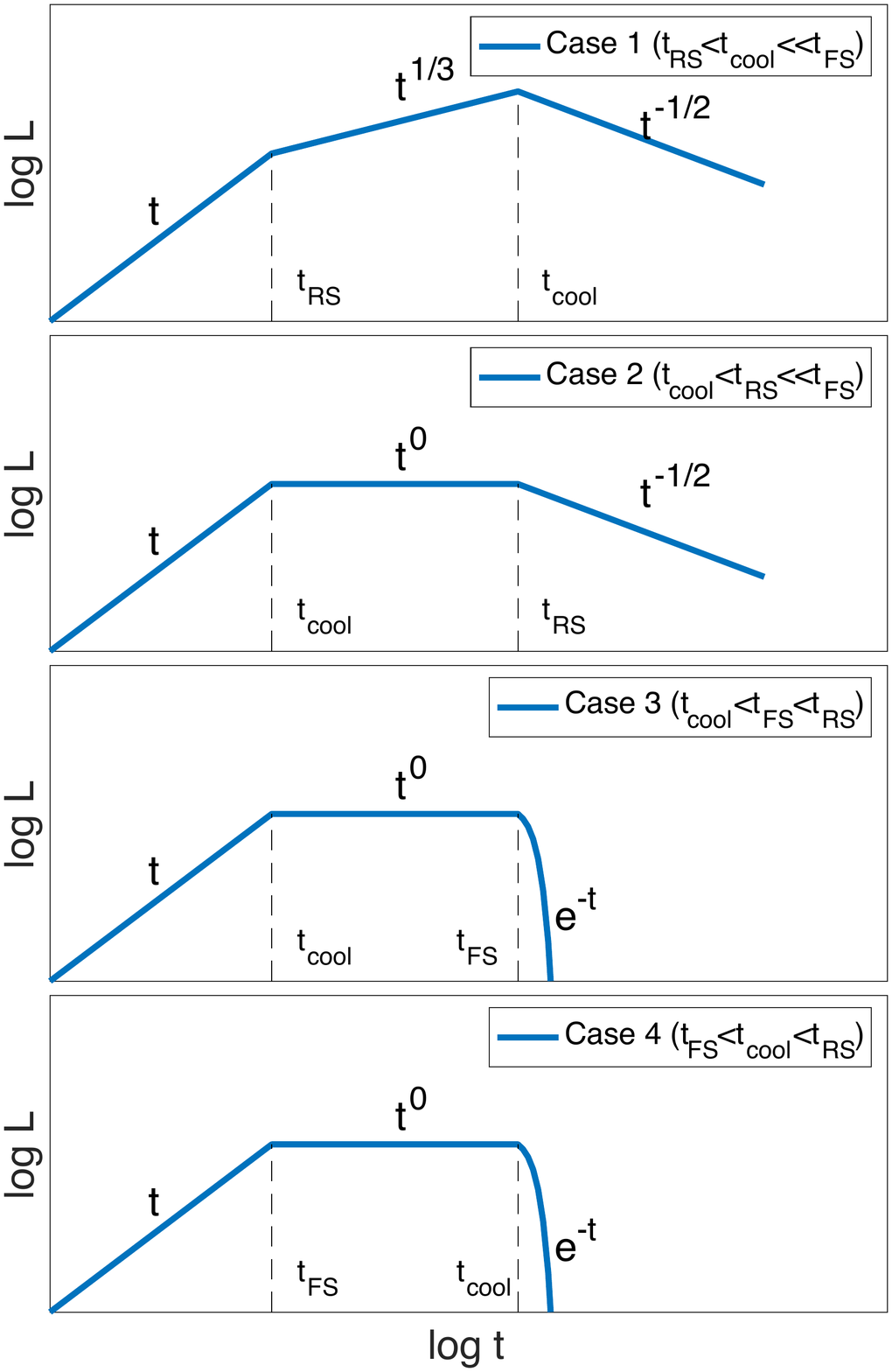}
\caption{ A sketch of the evolution of the bolometric luminosity of the interaction between the disk wind and the torus in different cases.\label{fig:cases}}
\end{figure}

We sketch the light curves of bolometric luminosity from the shocked torus in different cases  discussed above and we neglect the luminosity from the shocked wind because it is much smaller due to a low number density. The results are shown in Figure \ref{fig:cases}.

\subsection{Emission in X-ray} \label{subsec:X-ray}

Dynamical interactions between an ejected material and the ambient medium are thought to be essential in heating plasma to emit the X-rays \citep{1973SvA....16..749S, 1974ApJ...188..335M, 1975ApJ...197..621S, 1982ApJ...259..302C}. 
Electrons  can emit bremsstrahlung photons that provide the bulk of the thermal continuum \citep{1963asqu.book.....A,1964ApJ...140..470H}. 
The emissivity of bremsstrahlung in units of erg $\rm cm^{-3}$ $\rm s^{-1}$ $\rm Hz^{-1}$ is 
\citep{1979rpa..book.....R}:
\begin{equation}
\epsilon^{ff}_{\nu}=6.8\times 10^{-38}  T^{-\frac{1}{2}}e^{-\frac{h\nu}{kT}}\overline g_{ff}n_{e}\sum n_{i}Z^{2} ,
\end{equation}
where $\overline g_{ff}\approx 1$ is the gaunt-factor, $n_{e}, n_ {I} $ represent the number density of electrons and ions respectively.

For $t<t_{cool}$, the calculation of the bremsstrahlung emission can be approximated by assuming that the emitting mass is the swept-up mass and $n_{i}\approx 4n_{1}$ in the adiabatic phases \citep{1981ApJ...251..259C}.
Multiplying $\epsilon^{ff}_{\nu}$ by the volume of the shocked material, we have
\begin{align} \label{eqn:Lxray}
L_{\nu}&=V(t)\epsilon ^{ff}_{\nu}(t)
\end{align}
The X-ray luminosity $L_{X}$ can be obtained by integrating the above equation over the relevant frequency range.

In the radiative phase ($t>t_{cool}$), one can not use Equation (\ref{eqn:Lxray}) because $n_{i}\approx 4n_{1}$ is invalid in the non-adiabatic phase. However, as the temperature of the shocked plasma has become low,
 we adopt an exponential decay for $L_{X}(t)$ after $t=t_{cool}$.

\subsection{Emission in Radio} \label{subsec:Radio}
The shocked regions can also be a non-thermal synchrotron radio source. Shocks are believed to be responsible for accelerating particles \citep{1997ApJ...490..619S, 1998ApJ...492..219G, 2009A&ARv..17..409T} and the magnetic field required is generated as the shock sweeps up material \citep{1975MNRAS.171..263G}. Recent theories and observations have revealed that these shocks put some fraction of their energy into accelerated, non-thermal particles and magnetic field \citep{2008ARA&A..46...89R}. 

The synchrotron emission, from a power-law distribution of electrons concerning their Lorentz factor $dN/d\gamma \propto \gamma^{-p}$ at $\gamma > \gamma_{m}$, can be approximated by a broken power-law spectrum with three characteristic break frequencies \citep{1970ranp.book.....P,1998ApJ...497L..17S}. One is the self-absorption frequency, $\nu _{a}$, below which the system becomes optically thick. 
The second is $\nu _{m}$, the typical synchrotron frequency of electrons corresponding to the  minimal Lorentz factor $\gamma _{m}$.
The last one is $\nu_{c}$ which is taken into account because the electrons in a plasma emitting synchrotron radiation will be cooling down.

Assuming that a constant fraction $\epsilon_{e}$ of internal energy of the shocked medium goes into the non-thermal electrons and a fraction $\epsilon _{B}$ is converted into the magnetic field, $\nu_{m}$ is \citep{2013MNRAS.430.2121P}:
\begin{equation}\label{eqn:vm}
\nu _{m}=n^{\frac{1}{2}}_{1}\epsilon _{B,-1}^{\frac{1}{2}} \epsilon _{e,-1}^{2}\beta ^{5}\ \rm GHz,
\end{equation}
where $\beta=v_{1}/c$; $\epsilon _{B,-1}$ and $\epsilon _{e,-1}$ are  $\epsilon _{B}$ and $\epsilon _{e}$ normalized to 0.1.

The shock-generated magnetic field which is distributed throughout the shocked volume, and $\nu _{c}$ can be estimated as \citep{1998ApJ...497L..17S,2019MNRAS.487.4083Y}:
\begin{eqnarray}
B&=&(32 \pi \epsilon _{B}m_{H} n_{1})^{\frac{1}{2}} \beta c,\label{eqn:B}\\
\nu _{c}&=&10^{18}n^{-\frac{3}{2}}_{1}\epsilon _{B,-1}^{-\frac{3}{2}} \beta ^{-3}t^{-2}\ \rm GHz,
\end{eqnarray}
respectively. 

The synchrotron self-absorption frequency can be found by equating the absorption optical depth in the shocked region to unity:
\begin{equation}
\nu _{a}=D^{\frac{2}{p+4}}_{17} n^{\frac{6+p}{2(p+4)}}_{1}\epsilon _{B,-1}^{\frac{2+p}{2(p+4)}} \epsilon _{e,-1}^{\frac{2(p-1)}{p+4}}\beta ^{\frac{5p-2}{p+4}}\ \rm GHz.
\end{equation}
Where $D_{17}$ is the width of the shock region in units of $10^{17}$ cm.

\begin{figure}[htbp]
\centering
\includegraphics[width=8cm]{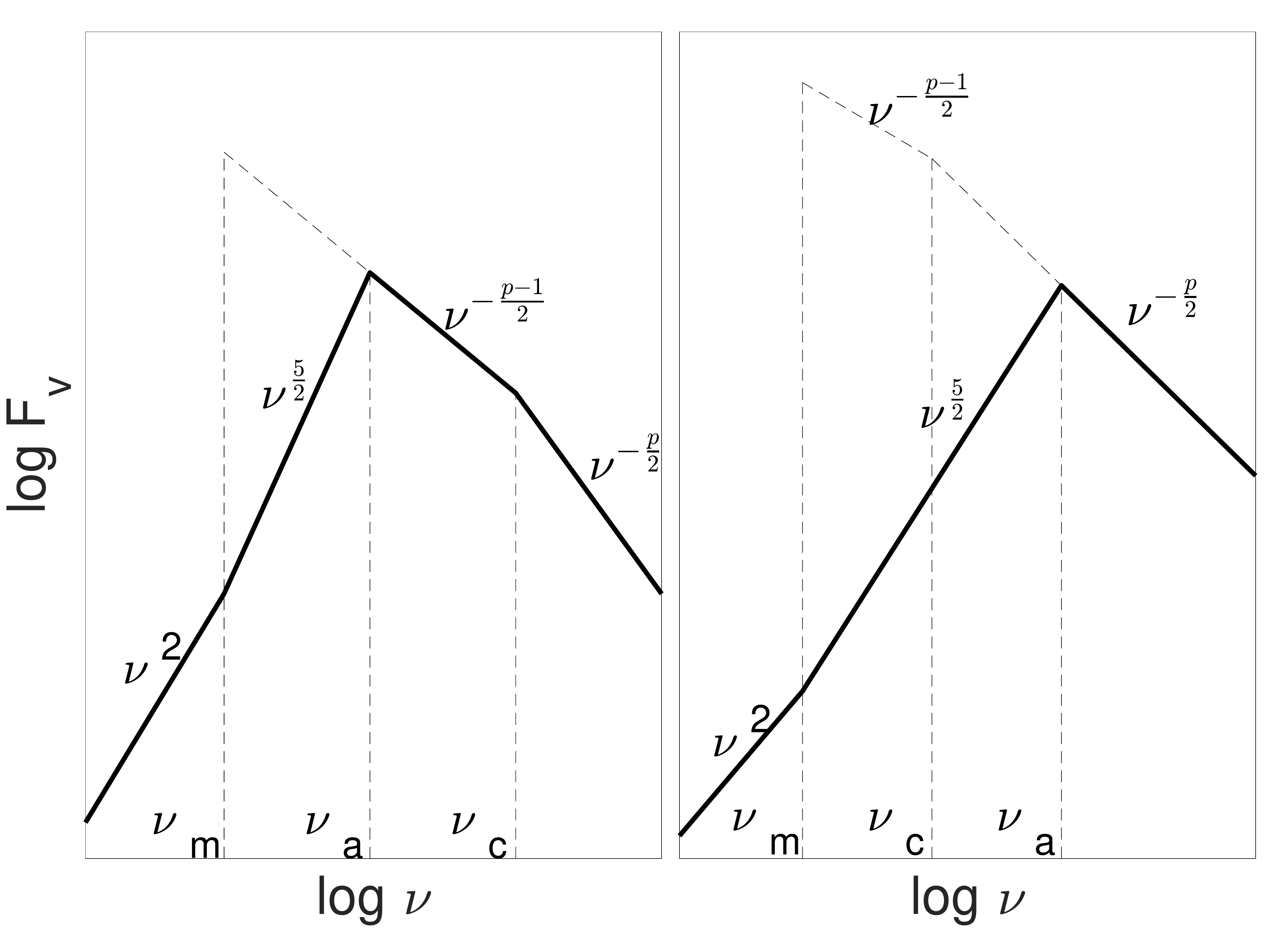}
\caption{A sketch of the synchrotron spectra of a shock in two cases.\label{fig:spectrum}}
\end{figure}

These three frequencies change over time  which will lead to the transformation of the synchrotron spectrum
from the weak self-absorption case to the strong self-absorption case once $\nu_{a}$ exceeds $\nu_{c}$ \citep{2013MNRAS.435.2520G}. 
Figure \ref{fig:spectrum} illustrates the two possible spectra, depending on the order of $\nu _{a}$, $\nu_{m}$ and $\nu_{c}$. 

For a randomly oriented electron velocity distribution, we can divide the radiation power of a single electron $P(\gamma_{e})=4 \sigma _{t}c\gamma^{2}_{e}B^{2}/24 \pi$ by the characteristic synchrotron frequency $\nu (\gamma_{e})=\gamma_{e}^{2}q_{e}B/2\pi m_{e}c$ to get the peak spectral power $P_{\nu,max}=\sigma _{t}m_{e}c^{2}B/3q_{e}$, which is independent of $\gamma_{e}$  \citep{1979rpa..book.....R,1998ApJ...497L..17S}. Summing all the electrons in the postshock fluid,  the unabsorbed synchrotron flux at $\nu_{m}$ can be obtained as :
\begin{equation} \label{eqn:Fvm}
F_{\nu_{m}}=\frac{V_{2}n_{2}P_{\nu,max}}{4\pi d^{2}}
\end{equation}
where $d$ is the distance to the source. 
The flux at any frequency can be found using these spectra and $F_{\nu_{m}}$.

\section{Application to PS1-10adi}
The evolution of the collision between wind and torus depends on its width to a great degree, because they would determine the order of $t_{FS}$, $t_{RS}$, and $t_{cool}$.

Before application, some quantities can be computed from observations.
The collision time $t_{in}$ is about 1,450 days after the main peak.
Under the circumstances that a typical velocity of the black hole disk wind is $v_{0}$=0.1 c \citep{2009MNRAS.400.2070S, 2017ApJ...843L..19M,2018ApJ...859L..20D}, 
the inner radius of the torus $R_{in}$ is approximately $0.1$ pc.
Besides, we can crudely estimate the total radiated energy by integrating the bolometric light curve between 1,500 and 2,000 rest-frame days which gives total energy of about $10^{50}$ erg, and that would be comparable to the kinetic energy of disk wind $E_{0}$.

\subsection{Cases of the wind and torus widths}
Below we will consider the wind-torus collision under different widths of wind and torus, and then examine whether they can produce results that are consistent with the observations of the late flare in PS1-10adi.
Finally, we would compare our analytic bolometric light curves to the observed ones of the rebrightening of PS1-10adi presented in \cite{2019ApJ...871...15J}.

\subsubsection{Narrow wind and wide torus }
Since the black hole can be in super-Eddington accretion for months, during which the disk wind can be launched due to the large radiation pressure, assuming that the disk wind has the same velocity $v_{0}$, then the width of the disk wind is  
 $W_{w}=$ $ v_{0}$ $\Delta t$, where we take $\Delta t =$ 30 days as an example. The density of the disk wind is thought to be 
\begin{align}
\rho_{w}(r,t+t_{in})&= \frac{M_{ej}}{\Delta t} \frac{1}{\Omega r^{2}v_{0}}
\end{align}

This density is almost homogenous while colliding with the cloud, from which  the mass density in front of the reverse shock can be considered unchanged all the time.

The wide torus means that the width of the torus can be considered infinitely wide. On this occasion, it takes about 25 days for the reverse shock to cross the wind.
We found results comparable to observations can be obtained with rational parameters in this case with $t_{RS}<t_{cool}\ll t_{FS}$, which corresponds to case 1, and we would analyze it in greater detail below in section 4.2.

\subsubsection{wide wind and wide torus}

When $W_{w}\approx R_{in}/2$ and the wind has a velocity spread of about order unity ($v_{0}/2 \sim v_{0}$), at any given time, we have
\begin{align}
\mathrm{d}M_{w}&=\Omega r^{2} \rho_{w}(r,t) \mathrm{d}r=\Omega v^{2}t^{3}\rho_{w}(r,t)\mathrm{d}v\\
M_{ej}&=\int \frac{\mathrm{d}M_{w}}{\mathrm{d}v} \mathrm{d}v
\end{align}
Considering $\mathrm{d}M_{w}/\mathrm{d}v \propto v^{a}$, we can obtain 
\begin{align}
\rho_{w}(r, t+t_{in})&=\frac{a+1}{1-2^{-(a+1)}}  \frac{M_{ej}}{\Omega r^{2} v_{0}^{a+1} (t+t_{in})}\\
v_{w}(r,t+t_{in})&=\frac{r}{t+t_{in}}
\end{align} 

Since the evolution timescale of the late flare in PS1-10adi is much smaller than $t_{in}$,
it is rational to assume that the wind density in front of the reverse shock is nearly unchanged and equals to the density at the moment of collision under the assumption that the wind mass density is uniform as:
\begin{align}
\rho_{w}(r,t+t_{in}) \approx \frac{M_{ej}}{\Omega R^{3}_{in}}
\end{align} 

Under this circumstance, it takes approximately 1,000 days for the reverse shock to cross the wind in line with equation (\ref{eqn:trs}). Results that are comparable to observations can not be obtained under this case according to the top two panels in Figure \ref{fig:cases} because it takes at least 1,000 days for $L_{B}$ to reach its peak while the observations show that this late flare reached its peak within only 100 days. 

\subsubsection{Wide wind and narrow torus}
Under the structural parameters setting in section 4.1.2, additionally, we consider that the width of the torus is so small that the forward shock would dissipate before $t_{RS}$. 
According to the bottom two panels in Figure \ref{fig:cases}, if we want to reproduce the features of the observed light curve of the late flare in PS1-10adi under this structure, we must require that both $t_{cool}$ and $t_{FS}$ are comparable to about 100 days.
According to equantion (\ref{eqn:tcool}) and (\ref{eqn:Lbol}),  we have 
\begin{align}\label{eqn:Ltcool}
L_{B}(t_{cool}) \approx 10^{40} n_{A}^{-\frac{1}{2}} n_{D}^{\frac{3}{2}} \ erg s^{-1}
\end{align}

Obviously, we can select the parameters to make sure that it would give a result comparable to observation. However we should bear in mind, while selecting the parameters, that $\alpha>1$ should be met and that the mass of the colliding disk wind with a width of $R_{in}/2$ should be in a reasonable range ($M_{ej}<1 \rm M_{\odot}$), as well as the number density of the torus.
In addition, we must check whether the cooling time is consistent with the observation under the selected parameters. And we found that the corresponding cooling times ($10^{9}-10^{12}$ s ) are much longer than the observation ($10^{7} $ s).

\subsection{Fits to the rebrightening of PS1-10adi}

\begin{figure} [htbp]
\centering
\includegraphics[width=7.5cm]{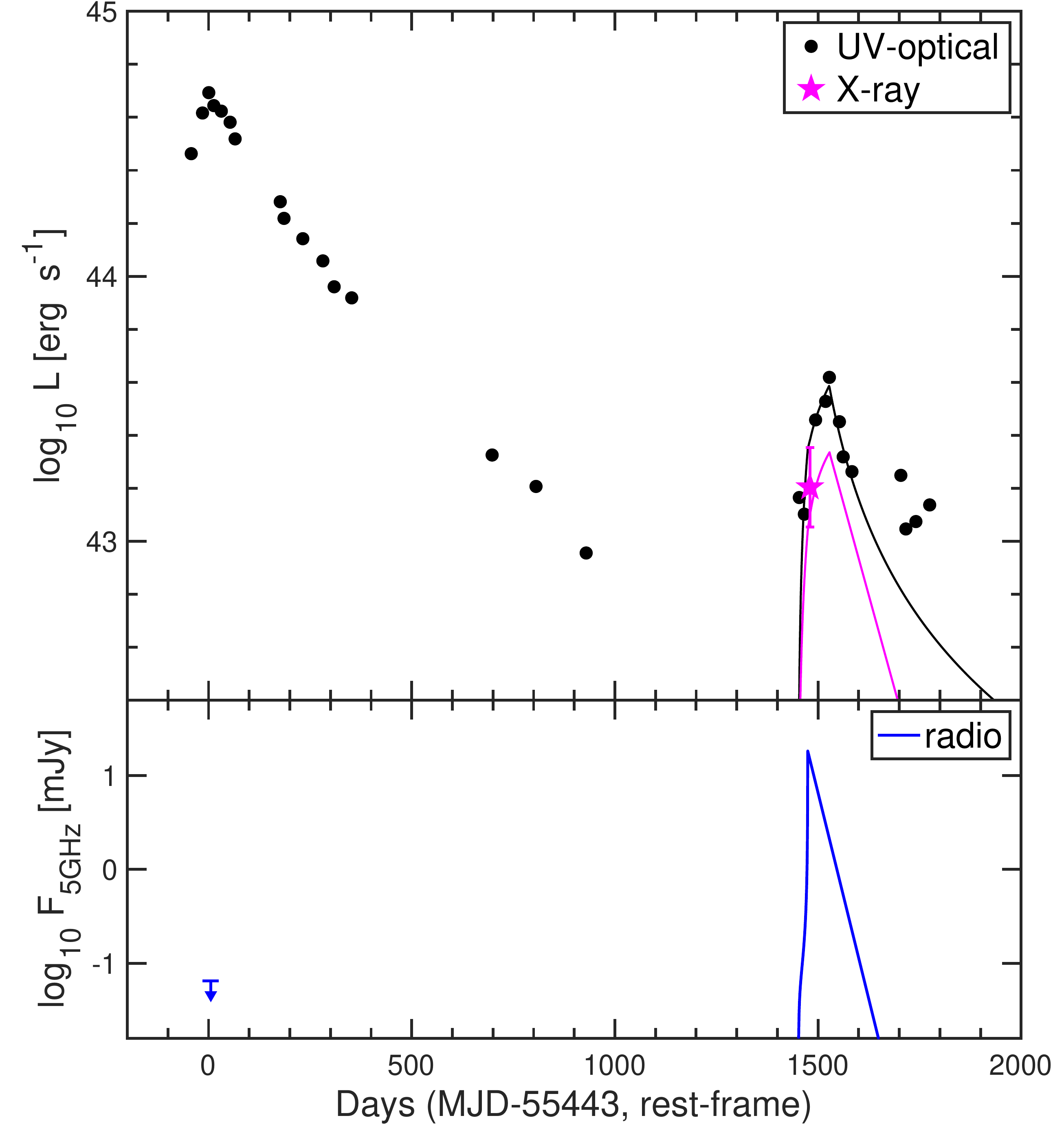} 
\caption{Fits of our model to observational data of PS1-10adi. Top panel: the black dots represent the UV-optical blackbody fitted luminosity,  and the X-ray detected during the rebrightening is plotted as a magenta five-pointed star \citep{2019ApJ...871...15J}. The solid black line is the bolometric light curve of our model while the solid magenta line is the luminosity of X-ray. Bottom panel:  the blue line is the flux of radio obtained with our model. The radio upper limit during the main peak, $F_{\rm 5 GHz} < 65$ $\rm \mu Jy$, is from \cite{2017NatAs...1..865K}. \label{fig:Lcalculated}}
\end{figure}

According to the  discussion above,  we think that the case where the wind is mainly distributed in a narrow width is a more reasonable possibility.
Based on this case, we compare our analytic bolometric light curves to the observed ones of the rebrightening of PS1-10adi presented in \cite{2019ApJ...871...15J} and shown in
Figure \ref{fig:Lcalculated}. We find that the features of the observed light curve are naturally reproduced by our model.

Here we have set the opening angle of the torus subtended to the black hole to be $\pi/6$, which gives a solid angle $\Omega \approx 3.3$. The covering factor $f_{c}=\Omega/4\pi$ determines the probability of disk wind to collide with torus \citep{2015ARA&A..53..365N, 2016MNRAS.458.3314C}. Since $f_{c}$ is of order unity, this wind-torus interaction is very likely to happen for TDEs in AGN.

On the spectral aspect, the observed UV-optical luminosity of the late flare in PS1-10adi is comparable to its X-ray luminosity, as is shown in Figure \ref{fig:Lcalculated}. However, in our model, the dominant radiation process is the thermal free-free emission, whose luminosity at the relevant temperature of $\sim 10^{7}$ K would come out predominantly in the X-rays, whereas the UV-optical luminosity would be only a small fraction of the bolometric one. This discrepancy might be accounted for by envisaging that the observed UV-optical emissions are from the reprocessing of most of the free-free emission by the surrounding low-density, low-temperature gas in the torus.  

In the best fitting case, the total kinetic energy of the disk wind is $E_{0}\approx 1.7\times10^{51}$ erg, the velocity of disk wind is  $v_{0}\approx0.08$ c,  the inner radius of torus is $R_{in}\approx 0.1$ pc and the gas number density of torus is $n_{0}\approx 3.5\times10^{7}\ \rm cm^{-3}$. These variables matching general values may give support to our model.
There seems to be a residual variability at about 1,700 days after the main peak. It is likely caused by the inhomogeneity in the torus or wind. 

\begin{figure}[htbp]
\centering
\includegraphics[width=8.5cm]{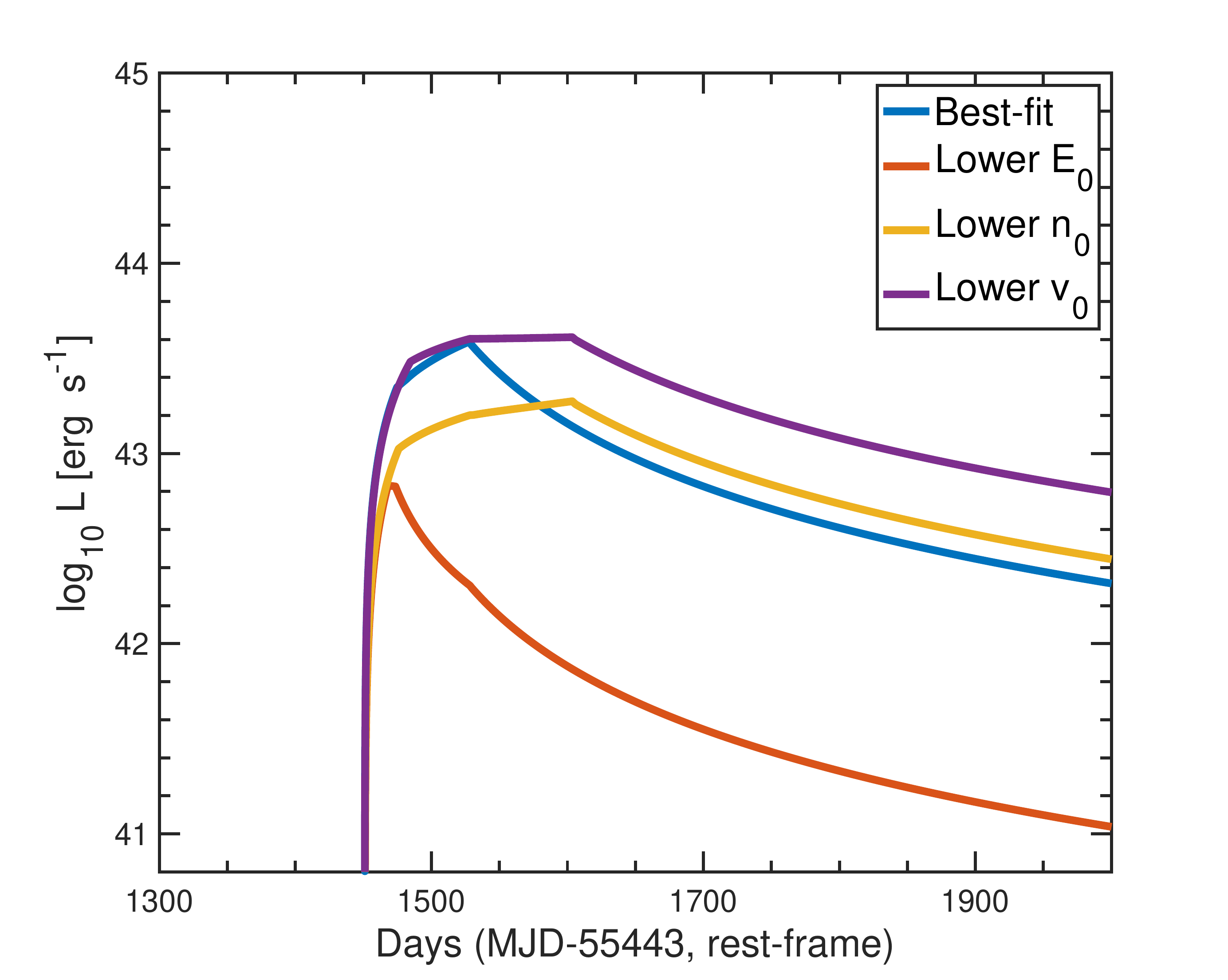}
\caption{Comparison of model light curves with different model parameters shown in Table \ref{table:aboutL}. \label{fig:models}}
\end{figure}

In order to check to see  if there is any degeneracy in the determination of various model parameter values, we vary those parameters each at a time, as were listed in Table \ref{table:aboutL}, to see how the bolometric luminosity light curve changes. The results are shown in figure \ref{fig:models}.
We conclude that the light curve mainly depends on $E_{0}$ and $n_{0}$. 
Any slight changes of these parameters may lead to diverse results. The higher $E_{0}$ will lead to a higher peak luminosity. The number density of the torus determines the timescale of the whole evolutionary process. The lower the number density is, the longer the timescale will be. Even if we increase the inner radius of the torus and the velocity of the disk wind to ensure that it would peak at about 1,500 days after the main peak, the resulted light curve would not be the same as the observed one for a higher velocity implies a lower mass and a shorter timescale of the whole evolution.
\begin{table}[htbp]
\centering
\caption{Varied model parameters with respect to the best-fit ones. (1) Model label, (2) Total energy of disk wind, (3) Torus density, (4) Wind speed. \label{table:aboutL}}
\setlength{\tabcolsep}{1.7mm} 
\begin{tabular}{cccc}
\toprule
 	  &	$E_{0}$    & 	$n_{0}$ 			& 	$v_{0}/c$		\\
Model &    (erg)	& (	$\rm cm^{-3}$) 	&   				\\
\midrule
Best-fit 			&  $1.7\times10^{51}$ 		& $3.5\times10^{7}$ 	& 0.08 \\
Lower $E_{0}$ 		&  $1.7\times10^{50}$ 		& $3.5\times10^{7}$ 	& 0.08 \\
Lower $n_{0}$ 		&  $1.7\times10^{51}$ 		& $1.5\times10^{7}$ 		& 0.08 \\
Lower $v_{0}$ 		&  $1.7\times10^{51}$ 		& $3.5\times10^{7}$ 	& 0.04   \\
\bottomrule  
\end{tabular}
\end{table}

We set $\epsilon _{e}=\epsilon _{b}=0.1$ and $p$ = 2.5 which are standard values for the synchrotron emission in fast shocks \citep{2006ApJ...651..381C, 2019MNRAS.487.4083Y}. 
The result is shown in Figure \ref{fig:Lcalculated} in blue line though the spectrum in self-strong absorption case that we adopt is simplified since we do not include the effects of strong absorption which would change the electron energy distribution according to \citep{ 2013MNRAS.435.2520G}. 
Contrary to the bolometric luminosity, the radio is mainly originated from the shocked wind because, comparing with the forward shock, the higher reverse shock velocity would lead to a stronger magnetic field and a higher characteristic frequency, even though the number density is small according to the equations (\ref{eqn:vm}-\ref{eqn:B}).

\section{Summary and Discussion} \label{sec:summary}
In this paper, we have developed a simple analytic model of the interaction between the disk wind and the torus for TDEs in AGN, which would produce strong shocks and a late flare. 
We also investigate the observational signatures and eventually obtain an analytic bolometric light curve model from the derived shock evolution.
This model can be used to fit observations to estimate the properties of torus and disk wind. 
Although our result was derived for a uniform medium, it should be approximately valid for the actual medium in torus which contains embedded clouds.
We have applied our bolometric light curve model to the rebrightening of PS1-10adi.
Close agreement with the observation is reached.
It shows that our general but somewhat approximate model, with reasonable choices of the above parameters, leads to a light curve consistent with observations.

In the analysis of the dynamics and radiation properties of the collision, we considered the impact of the widths of the disk wind and torus.
The diversity in widths of  the disk wind and torus would lead to the different order of $t_{RS}$, $t_{FS}$ and $t_{cool}$, which would  then make the evolution of luminosities completely disparate.
For $\alpha>1$ where the number density of the disk wind is lower than that of the torus, the luminosity is dominated by the shocked torus region for its higher number density.
But that is not the case for the non-thermal radio radiation, because it mainly depends on the shock velocity. Thus the radio emission is primarily from the shocked wind even with a lower number density.

In the application to the late flare in PS1-10adi, through the discussion of various cases, we found that the case, where the disk wind is distributed in a narrow width and the torus is wide, is a more reasonable possibility.
For other situation, either the evolution timescale is longer than observation, or results can't be obtained under rational parameters.

For the rebrightening of PS1-10adi, 
we find the following picture is consistent with the observation:
when a  mass of 0.3 $M_{\odot}$, which is distributed in a narrow width, is launched in the solid angle of the torus at a speed of 0.1 c, the peak luminosity of the late flare induced by the collision between disk wind and torus whose number density is about $10^{7}$ $\rm cm^{-3}$  would be $10^{43}$ erg $\rm s^{-1}$. 
Meanwhile, it may further radiate in X-ray and radio from the shocked medium.
As for the strong infrared emission which was also observed, we consider that it may be associated with dust heated in the torus \citep{1979ApJS...41..555H, 2019ApJ...871...15J}.

\cite{2017ApJ...843L..19M} have studied the interaction of wind from the AGN activity with a broad line region (BLR) for a separate signal. We do not attribute the TDE rebrightening discussed here to this interaction for the following two reasons. Firstly, BLRs are found to be located in the immediate vicinity of the central BHs in AGNs. If it was the interaction between the wind and the matter in BLRs, it would be difficult to explain why the signal did not appear until 1,500 days after and the luminosity increased at a very rapid rate. Secondly, the BLR is thought to be dust-free \citep{2015ARA&A..53..365N}, whereas PS1-10adi was detected in infrared in variation around the same time of the rebrightening \citep{2017NatAs...1..865K, 2019ApJ...871...15J}. Since an infrared signal is often associated with dust emission, the interaction with a dusty torus would be a more reasonable possibility.

Our model is set to be one-dimensional for simplicity. However, we should know that in reality, the interaction is not one-dimensional where the shocked matter would expand in the lateral direction, thereby avoiding collision with the densest part of the cloud, a two-dimensional situation studied numerically by  \cite{2020arXiv200910420M}.

\cite{2020arXiv200910420M} have analyzed the X-ray afterglow of PS1-10adi which is also attributed to the interaction between disk wind and clumpy clouds in the torus. 
According to their results, the torus density is about $10^{7}m_{H}$ $\rm cm^{-3}$ and the velocity of the disk wind is 0.11 c to which our results are similar.

The late flares of TDEs may be normal for TDEs in AGN and the late flares may also help us distinguish TDEs from other transient events.  
Future observations of similar events would help us better understand TDEs, as well as the gaseous and dusty torus around the central black hole in AGN by modeling them.
\\

We thank the anonymous referee for many constructive comments and suggestions that helped improve the quality of this paper.This work is supported by National Natural Science Foundation of China (12073091 and 11673078), Guangdong Basic and Applied Basic Research Foundation (2019A1515011119) and Guangdong Major Project of Basic and Applied Basic Research (2019B030302001).

\bibliography{bibfile}{}

\begin{thebibliography}{}
\expandafter\ifx\csname natexlab\endcsname\relax\def\natexlab#1{#1}\fi
\providecommand{\url}[1]{\href{#1}{#1}}
\providecommand{\dodoi}[1]{doi:~\href{http://doi.org/#1}{\nolinkurl{#1}}}
\providecommand{\doeprint}[1]{\href{http://ascl.net/#1}{\nolinkurl{http://ascl.net/#1}}}
\providecommand{\doarXiv}[1]{\href{https://arxiv.org/abs/#1}{\nolinkurl{https://arxiv.org/abs/#1}}}

\bibitem[{{Allen}(1963)}]{1963asqu.book.....A}
{Allen}, C.~W. 1963, {Astrophysical quantities}

\bibitem[{{Antonucci}(1993)}]{1993ARA&A..31..473A}
{Antonucci}, R. 1993, ARA$\&$A, 31, 473,
  \dodoi{10.1146/annurev.aa.31.090193.002353}

\bibitem[{{Blanchard} {et~al.}(2017){Blanchard}, {Nicholl}, {Berger},
  {Guillochon}, {Margutti}, {Chornock}, {Alexander}, {Leja}, \&
  {Drout}}]{2017ApJ...843..106B}
{Blanchard}, P.~K., {Nicholl}, M., {Berger}, E., {et~al.} 2017, ApJ, 843, 106,
  \dodoi{10.3847/1538-4357/aa77f7}

\bibitem[{{Blondin} {et~al.}(1998){Blondin}, {Wright}, {Borkowski}, \&
  {Reynolds}}]{1998ApJ...500..342B}
{Blondin}, J.~M., {Wright}, E.~B., {Borkowski}, K.~J., \& {Reynolds}, S.~P.
  1998, ApJ, 500, 342, \dodoi{10.1086/305708}

\bibitem[{{Chen} {et~al.}(2016){Chen}, {G{\'o}mez-Vargas}, \&
  {Guillochon}}]{2016MNRAS.458.3314C}
{Chen}, X., {G{\'o}mez-Vargas}, G.~A., \& {Guillochon}, J. 2016, MNRAS, 458,
  3314, \dodoi{10.1093/mnras/stw437}

\bibitem[{{Chevalier}(1974)}]{1974ApJ...188..501C}
{Chevalier}, R.~A. 1974, ApJ, 188, 501, \dodoi{10.1086/152740}

\bibitem[{{Chevalier}(1981)}]{1981ApJ...251..259C}
---. 1981, ApJ, 251, 259, \dodoi{10.1086/159460}

\bibitem[{{Chevalier}(1982)}]{1982ApJ...259..302C}
---. 1982, ApJ, 259, 302, \dodoi{10.1086/160167}

\bibitem[{{Chevalier} \& {Fransson}(2006)}]{2006ApJ...651..381C}
{Chevalier}, R.~A., \& {Fransson}, C. 2006, ApJ, 651, 381,
  \dodoi{10.1086/507606}

\bibitem[{{Christopher} {et~al.}(2005){Christopher}, {Scoville}, {Stolovy}, \&
  {Yun}}]{2005ApJ...622..346C}
{Christopher}, M.~H., {Scoville}, N.~Z., {Stolovy}, S.~R., \& {Yun}, M.~S.
  2005, ApJ, 622, 346, \dodoi{10.1086/427911}

\bibitem[{Cioffi \& McKee(1988)}]{cioffi_mckee_1988}
Cioffi, D., \& McKee, C. 1988, International Astronomical Union Colloquium,
  101, 435–438, \dodoi{10.1017/S0252921100102799}

\bibitem[{{Cioffi} {et~al.}(1988){Cioffi}, {McKee}, \&
  {Bertschinger}}]{1988ApJ...334..252C}
{Cioffi}, D.~F., {McKee}, C.~F., \& {Bertschinger}, E. 1988, ApJ, 334, 252,
  \dodoi{10.1086/166834}

\bibitem[{{Cox} \& {Daltabuit}(1971)}]{1971ApJ...167..113C}
{Cox}, D.~P., \& {Daltabuit}, E. 1971, ApJ, 167, 113, \dodoi{10.1086/151009}

\bibitem[{{Cox} \& {Tucker}(1969)}]{1969ApJ...157.1157C}
{Cox}, D.~P., \& {Tucker}, W.~H. 1969, ApJ, 157, 1157, \dodoi{10.1086/150144}

\bibitem[{{Dai} {et~al.}(2018){Dai}, {McKinney}, {Roth}, {Ramirez-Ruiz}, \&
  {Miller}}]{2018ApJ...859L..20D}
{Dai}, L., {McKinney}, J.~C., {Roth}, N., {Ramirez-Ruiz}, E., \& {Miller},
  M.~C. 2018, ApJL, 859, L20, \dodoi{10.3847/2041-8213/aab429}

\bibitem[{{Gaisser} {et~al.}(1998){Gaisser}, {Protheroe}, \&
  {Stanev}}]{1998ApJ...492..219G}
{Gaisser}, T.~K., {Protheroe}, R.~J., \& {Stanev}, T. 1998, ApJ, 492, 219,
  \dodoi{10.1086/305011}

\bibitem[{{Gao} {et~al.}(2013){Gao}, {Lei}, {Wu}, \&
  {Zhang}}]{2013MNRAS.435.2520G}
{Gao}, H., {Lei}, W.-H., {Wu}, X.-F., \& {Zhang}, B. 2013, MNRAS, 435, 2520,
  \dodoi{10.1093/mnras/stt1461}

\bibitem[{{Guillochon} {et~al.}(2016){Guillochon}, {McCourt}, {Chen},
  {Johnson}, \& {Berger}}]{2016ApJ...822...48G}
{Guillochon}, J., {McCourt}, M., {Chen}, X., {Johnson}, M.~D., \& {Berger}, E.
  2016, ApJ, 822, 48, \dodoi{10.3847/0004-637X/822/1/48}

\bibitem[{{Gull}(1973)}]{1973MNRAS.161...47G}
{Gull}, S.~F. 1973, MNRAS, 161, 47, \dodoi{10.1093/mnras/161.1.47}

\bibitem[{{Gull}(1975)}]{1975MNRAS.171..263G}
---. 1975, MNRAS, 171, 263, \dodoi{10.1093/mnras/171.2.263}

\bibitem[{{Hamilton} \& {Sarazin}(1984)}]{1984ApJ...281..682H}
{Hamilton}, A.~J.~S., \& {Sarazin}, C.~L. 1984, ApJ, 281, 682,
  \dodoi{10.1086/162145}

\bibitem[{{Heiles}(1964)}]{1964ApJ...140..470H}
{Heiles}, C. 1964, ApJ, 140, 470, \dodoi{10.1086/147941}

\bibitem[{{Hollenbach} \& {McKee}(1979)}]{1979ApJS...41..555H}
{Hollenbach}, D., \& {McKee}, C.~F. 1979, ApJS, 41, 555, \dodoi{10.1086/190631}

\bibitem[{{Jackson} {et~al.}(1993){Jackson}, {Geis}, {Genzel}, {Harris},
  {Madden}, {Poglitsch}, {Stacey}, \& {Townes}}]{1993ApJ...402..173J}
{Jackson}, J.~M., {Geis}, N., {Genzel}, R., {et~al.} 1993, ApJ, 402, 173,
  \dodoi{10.1086/172120}

\bibitem[{{Jiang} {et~al.}(2019){Jiang}, {Wang}, {Mou}, {Liu}, {Dou}, {Sheng},
  \& {Wang}}]{2019ApJ...871...15J}
{Jiang}, N., {Wang}, T., {Mou}, G., {et~al.} 2019, ApJ, 871, 15,
  \dodoi{10.3847/1538-4357/aaf6b2}

\bibitem[{{Jiang} {et~al.}(2016){Jiang}, {Guillochon}, \&
  {Loeb}}]{2016ApJ...830..125J}
{Jiang}, Y.-F., {Guillochon}, J., \& {Loeb}, A. 2016, ApJ, 830, 125,
  \dodoi{10.3847/0004-637X/830/2/125}

\bibitem[{{Jim{\'e}nez} {et~al.}(2019){Jim{\'e}nez}, {Tenorio-Tagle}, \&
  {Silich}}]{2019MNRAS.488..978J}
{Jim{\'e}nez}, S., {Tenorio-Tagle}, G., \& {Silich}, S. 2019, MNRAS, 488, 978,
  \dodoi{10.1093/mnras/stz1749}

\bibitem[{{Kankare} {et~al.}(2017){Kankare}, {Kotak}, {Mattila}, {Lundqvist},
  {Ward}, {Fraser}, {Lawrence}, {Smartt}, {Meikle}, {Bruce}, {Harmanen},
  {Hutton}, {Inserra}, {Kangas}, {Pastorello}, {Reynolds},
  {Romero-Ca{\~n}izales}, {Smith}, {Valenti}, {Chambers}, {Hodapp}, {Huber},
  {Kaiser}, {Kudritzki}, {Magnier}, {Tonry}, {Wainscoat}, \&
  {Waters}}]{2017NatAs...1..865K}
{Kankare}, E., {Kotak}, R., {Mattila}, S., {et~al.} 2017, Nature Astronomy, 1,
  865, \dodoi{10.1038/s41550-017-0290-2}

\bibitem[{{Karzas} \& {Latter}(1961)}]{1961ApJS....6..167K}
{Karzas}, W.~J., \& {Latter}, R. 1961, ApJS, 6, 167, \dodoi{10.1086/190063}

\bibitem[{{Komossa} {et~al.}(2008){Komossa}, {Zhou}, {Wang}, {Ajello}, {Ge},
  {Greiner}, {Lu}, {Salvato}, {Saxton}, {Shan}, {Xu}, \&
  {Yuan}}]{2008ApJ...678L..13K}
{Komossa}, S., {Zhou}, H., {Wang}, T., {et~al.} 2008, ApJL, 678, L13,
  \dodoi{10.1086/588281}

\bibitem[{{Krolik} \& {Begelman}(1988)}]{1988ApJ...329..702K}
{Krolik}, J.~H., \& {Begelman}, M.~C. 1988, ApJ, 329, 702,
  \dodoi{10.1086/166414}

\bibitem[{{Landau} \& {Lifshitz}(1959)}]{1959flme.book.....L}
{Landau}, L.~D., \& {Lifshitz}, E.~M. 1959, {Fluid mechanics}

\bibitem[{{Lu} \& {Bonnerot}(2020)}]{2020MNRAS.492..686L}
{Lu}, W., \& {Bonnerot}, C. 2020, MNRAS, 492, 686,
  \dodoi{10.1093/mnras/stz3405}

\bibitem[{{Lu} {et~al.}(2016){Lu}, {Kumar}, \& {Evans}}]{2016MNRAS.458..575L}
{Lu}, W., {Kumar}, P., \& {Evans}, N.~J. 2016, MNRAS, 458, 575,
  \dodoi{10.1093/mnras/stw307}

\bibitem[{{McKee}(1974)}]{1974ApJ...188..335M}
{McKee}, C.~F. 1974, ApJ, 188, 335, \dodoi{10.1086/152721}

\bibitem[{{McKee} \& {Hollenbach}(1980)}]{1980ARA&A..18..219M}
{McKee}, C.~F., \& {Hollenbach}, D.~J. 1980, ARA$\&$A, 18, 219,
  \dodoi{10.1146/annurev.aa.18.090180.001251}

\bibitem[{{Mezger} {et~al.}(1996){Mezger}, {Duschl}, \&
  {Zylka}}]{1996A&ARv...7..289M}
{Mezger}, P.~G., {Duschl}, W.~J., \& {Zylka}, R. 1996, A$\&$A Rv, 7, 289,
  \dodoi{10.1007/s001590050007}

\bibitem[{{Moriya} {et~al.}(2017){Moriya}, {Tanaka}, {Morokuma}, \&
  {Ohsuga}}]{2017ApJ...843L..19M}
{Moriya}, T.~J., {Tanaka}, M., {Morokuma}, T., \& {Ohsuga}, K. 2017, ApJL, 843,
  L19, \dodoi{10.3847/2041-8213/aa7af3}

\bibitem[{{Mou} {et~al.}(2020){Mou}, {Dou}, {Jiang}, {Wang}, {Guo}, {Wang},
  {Wang}, {Shu}, {He}, {Zhang}, \& {Sun}}]{2020arXiv200910420M}
{Mou}, G., {Dou}, L., {Jiang}, N., {et~al.} 2020, arXiv e-prints,
  arXiv:2009.10420.
\newblock \doarXiv{2009.10420}

\bibitem[{{Netzer}(2015)}]{2015ARA&A..53..365N}
{Netzer}, H. 2015, ARA$\&$A, 53, 365,
  \dodoi{10.1146/annurev-astro-082214-122302}

\bibitem[{{Ostriker} \& {McKee}(1988)}]{1988RvMP...60....1O}
{Ostriker}, J.~P., \& {McKee}, C.~F. 1988, Reviews of Modern Physics, 60, 1,
  \dodoi{10.1103/RevModPhys.60.1}

\bibitem[{{Pacholczyk}(1970)}]{1970ranp.book.....P}
{Pacholczyk}, A.~G. 1970, {Radio astrophysics. Nonthermal processes in galactic
  and extragalactic sources}

\bibitem[{{Phinney}(1989)}]{1989IAUS..136..543P}
{Phinney}, E.~S. 1989, in The Center of the Galaxy, ed. M.~{Morris}, Vol. 136,
  543

\bibitem[{{Piran} {et~al.}(2013){Piran}, {Nakar}, \&
  {Rosswog}}]{2013MNRAS.430.2121P}
{Piran}, T., {Nakar}, E., \& {Rosswog}, S. 2013, MNRAS, 430, 2121,
  \dodoi{10.1093/mnras/stt037}

\bibitem[{{Raymond} {et~al.}(1976){Raymond}, {Cox}, \&
  {Smith}}]{1976ApJ...204..290R}
{Raymond}, J.~C., {Cox}, D.~P., \& {Smith}, B.~W. 1976, ApJ, 204, 290,
  \dodoi{10.1086/154170}

\bibitem[{{Rees}(1988)}]{1988Natur.333..523R}
{Rees}, M.~J. 1988, Nature, 333, 523, \dodoi{10.1038/333523a0}

\bibitem[{{Reynolds}(2008)}]{2008ARA&A..46...89R}
{Reynolds}, S.~P. 2008, ARA$\&$A, 46, 89,
  \dodoi{10.1146/annurev.astro.46.060407.145237}

\bibitem[{{Reynolds}(2017)}]{2017hsn..book.1981R}
---. 2017, {Dynamical Evolution and Radiative Processes of Supernova Remnants},
  ed. A.~W. {Alsabti} \& P.~{Murdin}, 1981,
  \dodoi{10.1007/978-3-319-21846-5_89}

\bibitem[{{Rybicki} \& {Lightman}(1979)}]{1979rpa..book.....R}
{Rybicki}, G.~B., \& {Lightman}, A.~P. 1979, {Radiative processes in
  astrophysics}

\bibitem[{{Sadowski} {et~al.}(2016){Sadowski}, {Tejeda}, {Gafton}, {Rosswog},
  \& {Abarca}}]{2016MNRAS.458.4250S}
{Sadowski}, A., {Tejeda}, E., {Gafton}, E., {Rosswog}, S., \& {Abarca}, D.
  2016, MNRAS, 458, 4250, \dodoi{10.1093/mnras/stw589}

\bibitem[{{Sari} {et~al.}(1998){Sari}, {Piran}, \&
  {Narayan}}]{1998ApJ...497L..17S}
{Sari}, R., {Piran}, T., \& {Narayan}, R. 1998, ApJL, 497, L17,
  \dodoi{10.1086/311269}

\bibitem[{{Sedov}(1959)}]{1959sdmm.book.....S}
{Sedov}, L.~I. 1959, {Similarity and Dimensional Methods in Mechanics}

\bibitem[{{Sgro}(1975)}]{1975ApJ...197..621S}
{Sgro}, A.~G. 1975, ApJ, 197, 621, \dodoi{10.1086/153552}

\bibitem[{{Shklovskii}(1973)}]{1973SvA....16..749S}
{Shklovskii}, I.~S. 1973, Soviet Ast., 16, 749

\bibitem[{{Shklovsky}(1968)}]{1968supe.book.....S}
{Shklovsky}, J.~S. 1968, {Supernovae}

\bibitem[{{Shull} \& {McKee}(1979)}]{1979ApJ...227..131S}
{Shull}, J.~M., \& {McKee}, C.~F. 1979, ApJ, 227, 131, \dodoi{10.1086/156712}

\bibitem[{{Silich} \& {Tenorio-Tagle}(2018)}]{2018MNRAS.478.5112S}
{Silich}, S., \& {Tenorio-Tagle}, G. 2018, MNRAS, 478, 5112,
  \dodoi{10.1093/mnras/sty1383}

\bibitem[{{Strubbe} \& {Quataert}(2009)}]{2009MNRAS.400.2070S}
{Strubbe}, L.~E., \& {Quataert}, E. 2009, MNRAS, 400, 2070,
  \dodoi{10.1111/j.1365-2966.2009.15599.x}

\bibitem[{{Sturner} {et~al.}(1997){Sturner}, {Skibo}, {Dermer}, \&
  {Mattox}}]{1997ApJ...490..619S}
{Sturner}, S.~J., {Skibo}, J.~G., {Dermer}, C.~D., \& {Mattox}, J.~R. 1997,
  ApJ, 490, 619, \dodoi{10.1086/304894}

\bibitem[{{Taylor}(1950)}]{1950RSPSA.201..159T}
{Taylor}, G. 1950, Proceedings of the Royal Society of London Series A, 201,
  159, \dodoi{10.1098/rspa.1950.0049}

\bibitem[{{Treumann}(2009)}]{2009A&ARv..17..409T}
{Treumann}, R.~A. 2009, A$\&$A Rv, 17, 409, \dodoi{10.1007/s00159-009-0024-2}

\bibitem[{{Truelove} \& {McKee}(1999)}]{1999ApJS..120..299T}
{Truelove}, J.~K., \& {McKee}, C.~F. 1999, ApJS, 120, 299,
  \dodoi{10.1086/313176}

\bibitem[{{Vink}(2012)}]{2012A&ARv..20...49V}
{Vink}, J. 2012, A$\&$A Rv, 20, 49, \dodoi{10.1007/s00159-011-0049-1}

\bibitem[{{Yalinewich} {et~al.}(2019){Yalinewich}, {Steinberg}, {Piran}, \&
  {Krolik}}]{2019MNRAS.487.4083Y}
{Yalinewich}, A., {Steinberg}, E., {Piran}, T., \& {Krolik}, J.~H. 2019, MNRAS,
  487, 4083, \dodoi{10.1093/mnras/stz1567}

\end{thebibliography}

\end{document}